\documentclass[twoside,showpacs,superscriptaddress,twocolumn,floatfix,a4paper,aps,pra]{revtex4}

\usepackage{color}
\usepackage{graphicx}
\usepackage[utf8x]{inputenc}
\usepackage{hyperref}

\newcommand{\etal}{\textit{et al. }}

\begin{document}

\title{Knowing ``where the photons have been''}

\author{Karol Bartkiewicz} \email{bark@amu.edu.pl}
\affiliation{Faculty of Physics, Adam Mickiewicz University,
PL-61-614 Pozna\'n, Poland}
\affiliation{RCPTM, Joint Laboratory of Optics of Palacký University and Institute of Physics of Academy of Sciences of the Czech Republic, 17. listopadu 12, 771 46 Olomouc, Czech Republic}

\author{Antonín Černoch} \email{antonin.cernoch@upol.cz}
\affiliation{RCPTM, Joint Laboratory of Optics of Palacký University and Institute of Physics of Academy of Sciences of the Czech Republic, 17. listopadu 12, 771 46 Olomouc, Czech Republic}

\author{Dalibor Javůrek}
\affiliation{RCPTM, Joint Laboratory of Optics of Palacký University and Institute of Physics of Academy of Sciences of the Czech Republic, 17. listopadu 12, 771 46 Olomouc, Czech Republic}

\author{Karel Lemr} \email{k.lemr@upol.cz}
\affiliation{RCPTM, Joint Laboratory of Optics of Palacký University and Institute of Physics of Academy of Sciences of the Czech Republic, 17. listopadu 12, 771 46 Olomouc, Czech Republic}

\author{Jan Soubusta}
\affiliation{Institute of Physics of Academy of Sciences of the Czech Republic, Joint Laboratory of Optics of PU and IP AS CR, 
   17. listopadu 50A, 772 07 Olomouc, Czech Republic}

\author{Jiří Svozilík}
\affiliation{RCPTM, Joint Laboratory of Optics of Palacký University and Institute of Physics of Academy of Sciences of the Czech Republic, 17. listopadu 12, 771 46 Olomouc, Czech Republic}

\date{\today}

\begin{abstract}
Linear-optical interferometers play a key role in designing circuits for quantum 
information processing and quantum communications. Even though nested Mach-Zehnder 
interferometers appear easy to describe, there are occasions when they provide 
unintuitive results. This paper explains the results of {a} highly discussed experiment performed by Danan \etal  [Phys. Rev. Lett. {\bf 111}, 240402 (2013)] using a standard approach. We 
provide a simple and intuitive one-state vector formalism capable of interpreting their 
experiment. Additionally, we cross-check{ed} our model with a classical-physics based 
approach and found that both models are in complete agreement. We argue that the quantity used in the mentioned experiment is not a suitable which-path witness producing seemingly contra-intuitive results. To circumvent this issue, we establish a more reliable which-path witness and show that it yields well expected outcomes of the experiment.
\end{abstract}


\maketitle

In quantum mechanics (QM) particles are assigned a wave function used to 
describe their properties \cite{Feynman2011}. This approach sometimes leads to 
conclusions about experimental results {that seem to contradict intuitive estimations
based on classical physics~\cite{Einstein1935,Blaylock2010}}.
Despite this flaw, QM is currently widely accepted as  a theory 
\cite{Jammer1966,Feynman2011}, that makes
accurate predictions in agreement with the performed experiments.
Therefore, it is considered valid regularly used for the interpretation of 
the results of the corresponding experiments.

\begin{figure}[!h!]
\includegraphics[width=\columnwidth]{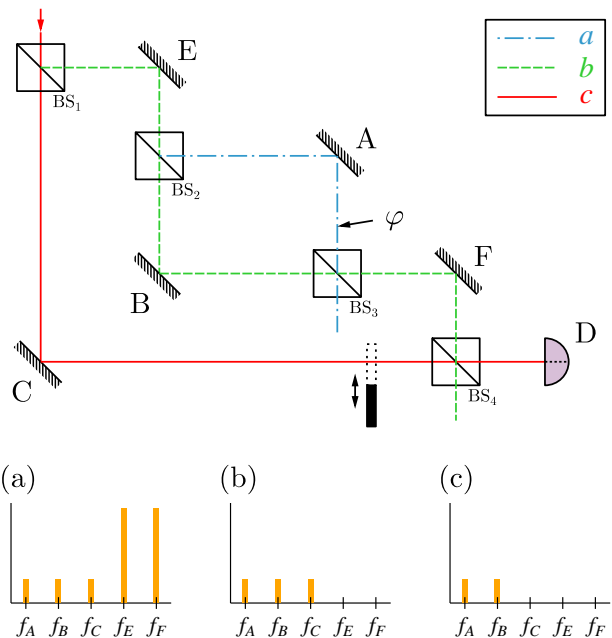}
\caption{\label{fig} (Color online) {Drawing} of the experimental setup with two 
nested Mach-Zender interferometers. (a) Power spectrum on detector D shows all 
frequencies of mirror oscillations {for} the phase $\varphi = \pi$, (b) power 
spectra {for} $\varphi = 0$, (c) power spectrum for phase  $\varphi = 0$ and 
the lower path $\hat{c}$ blocked as predicted by our approach.}
\end{figure}

Recently, an experiment {that contained} counter-intuitive features was proposed and 
realized by Danan \etal \cite{Danan2013}. The authors used nested Mach-Zehnder 
interferometers (MZI), shown in Fig.~\ref{fig}, {and} mirrors (A, B, C, E, F) vibrating 
with different frequencies, in order to leave a mark on {}passing photons. 
At one selected output port of the interferometer, the photons were detected 
by a quad-cell detector D {capable of tracing} the spatial vibrations of the photon 
beam. After measurement, the collected signal was further processed and subjected to 
{the} Fourier transform. From the obtained frequencies of vibrations, the authors judged 
whether the detected photons {have interacted with} the mirror {that was} 
oscillating at this particular frequency.

The results {described} in the article {by Danan \etal \cite{Danan2013}} 
were interpreted by means of two-state vector formalism (TSVF)
and weak values. {Both the results and their unusual} interpretation {were questioned}
\cite{Wiesniak2014,Salih2014,Svensson2014,Huang2014,Li2014}. 
{The critical comments pertained to the } visibility 
of interference inside the interferometer, {the} correct application of TSVF, 
{the} processing of the obtained data and {its} validity. 
{Until now, no-one has managed to provide the}
theoretical {calculations} and {the} interpretation of the experimental 
results {using only} the standard one-state vector quantum-mechanical approach. 
{Assuming the above-mentioned approach makes it possible to interpret
the results from Ref.~\cite{Danan2013} and verify their congruity with QM.
Consequently, we would manage to shed more light on the} ongoing
discussion {regarding the legitimacy of the experimental data and its
interpretation. In our opinion, this would be useful for describing} 
experiments similar to \cite{Danan2013} {clarifying} the debate about
the experiment.

In this paper, we present {a relevant} standard one-state vector formalism
and {describe the}  evolution of 
the state vector {as it passes} the MZI {in the direction of} the detector. 
We describe the post selection process and derive the probabilities of 
{detecting} at particular frequencies ({that correspond to the vibrations
of the} individual mirrors). {The} obtained results are compared with 
the experimental data presented by Danan \etal \cite{Danan2013}.
{Finally, we use classical optics to describe the transverse profiles of the 
light beams used in the experiment and we apply the result to
validate the one-state vector outcomes at the classical limit.}

{A correct description of} Danan \etal experiment
{needs to consider all the} photon modes present in the setup.
Apart from the spatial modes, {additional modes are} introduced 
by vibrations of the mirrors ({referred as the} ``frequency 
modes''). {The modes must} be taken into account, since they {differentiate},
at least in principle, {between} the respective paths of the photon. 
{Our} analysis {uses} the formalism of annihilation operators and 
their transformations on beam splitters (BS). {Spatial} modes are 
labeled by operators $\hat{a}$, $\hat{b}$ and $\hat{c}$, while the 
frequency modes are marked by {binary numbers}. {The} five frequency modes {are} 
introduced by the mirrors A, B, C, E or F. {They are marked by 
consecutive binary indices after the symbol naming the spatial mode. }
{The index value is either 0 or 1, indicating whether mode frequency
was modulated by a corresponding mirror (1) or not (0).}
So{,} for instance{,} if spatial mode $\hat{a}$ was frequency modulated by mirrors A, E and F, it 
would be indexed as $\hat{a}_{10011}$.

{At the beginning} the photon {is} in the mode $\hat{c}_{00000}$ 
(see setup depicted in Fig.~\ref{fig}). 
The first beam-splitter (BS$_1$) divides the beam with {an}
intensity ratio {of} 1:2. {As a result,} the spatial mode gets 
transformed to
\begin{equation}
  \hat{c}_{00000} \rightarrow \frac{1}{\sqrt{3}} \hat{c}_{00000} 
  + \sqrt{\frac{2}{3}}\mathrm{i} \hat{b}_{00000}.
\end{equation}
In the outer arm, the spatial mode $\hat{c}$ {interacts with the} vibrating mirror C, while 
the reflected mode $\hat{b}$ {comes into contact with} the mirror E. 
{This} is described by transition
\begin{equation}
  \frac{1}{\sqrt{3}} \hat{c}_{00000} + \sqrt{\frac{2}{3}}\mathrm{i} \hat{b}_{00000} 
  \rightarrow 
  \frac{1}{\sqrt{3}} \hat{c}_{00100} + \sqrt{\frac{2}{3}}\mathrm{i} \hat{b}_{00010}.
\end{equation}

The spatial mode $\hat{b}$ now enters the inner interferometer formed by two balanced 
beam splitters. {The} first beam splitter, BS$_2$, {transforms it} to
\begin{equation}
  \hat{b}_{00010} \rightarrow 
  \frac{1}{\sqrt{2}} \hat{b}_{00010} + \frac{1}{\sqrt{2}}i \hat{a}_{00010}.
\end{equation}
{The v}ibrating mirrors A and B then have the following effect
\begin{equation}
  \frac{1}{\sqrt{2}}\left( \hat{b}_{00010} + i \hat{a}_{00010} \right)
  \rightarrow 
  \frac{1}{\sqrt{2}} \left(\hat{b}_{01010} + i \hat{a}_{10010}\right).
\end{equation}
{The d}ifference between {the} lengths of {the} upper and lower arms of 
the inner interferometer introduce{s an} additional phase shift that {can be
attributed}  solely to the mode $\hat{a}$
\begin{equation}
  \hat{a}_{10010} \rightarrow 
  \mathrm{e}^{\mathrm{i}\varphi}\hat{a}_{10010}.
\end{equation}

{The} modes $\hat{b}$ and $\hat{a}$ get recombined {on} the second 
beam splitter of the inner interferometer{, i.e.,} BS$_3$.
{At this point we disregard the outgoing mode $\hat{a}$}
since only the mode $\hat{b}$ can further contribute to {photon}
detection, {hence}  
\begin{equation}
  \frac{1}{\sqrt{2}} \left(\hat{b}_{01010} 
    + \mathrm{i} \mathrm{e}^{\mathrm{i}\varphi}\hat{a}_{10010}\right) 
  \rightarrow 
  \frac{1}{2}\left(\hat{b}_{01010} 
    - \mathrm{e}^{\mathrm{i}\varphi}\hat{b}_{10010}\right).
\end{equation}
{The} output mode of the inner interferometer{, $\hat{b}$,} 
meets the last vibrating mirror{, F,}
\begin{equation}
 \frac{1}{2}\left(\hat{b}_{01010} - \mathrm{e}^{\mathrm{i}\varphi}\hat{b}_{10010}\right) 
 \rightarrow 
 \frac{1}{2}\left(\hat{b}_{01011} - \mathrm{e}^{\mathrm{i}\varphi}\hat{b}_{10011}\right).
\end{equation}
{F}inally, we recombine the modes $\hat{c}$ and $\hat{b}$ on the last unbalanced beam 
splitter BS$_4$ ({identical to} BS$_1$). The final form of the spatial mode $\hat{c}$, 
{describing photons that reach} detector D, is
\begin{equation}
  \frac{1}{3}\left(\hat{c}_{00100} - \hat{c}_{01011} 
    + \mathrm{e}^{\mathrm{i}\varphi} \hat{c}_{10011}\right).
\end{equation}

{As in the case of the} inner interferometer, we discard the output mode 
$\hat{b}$ that can not lead to photon detection at the detector D.
We assume {that the photons entering this setup are} single photon{s}
that can be described in therm of the creation operator $\hat{a}^\dagger|0\rangle$.
The output state in {the} Fock basis then reads
\begin{eqnarray}\label{eq:finalFock}
  |\psi_\mathrm{out}\rangle 
  &=& \frac{1}{3} \left( \hat{a}^\dagger_{00100}-\hat{a}^\dagger_{01011}
                         +\mathrm{e}^{\mathrm{i}\varphi} \hat{a}^\dagger_{10011}
                    \right)|0\rangle \nonumber\\
  &=& \frac{1}{3}\left(|1\rangle_{00100} - |1\rangle_{01011} 
                       + \mathrm{e}^{\mathrm{i}\varphi}|1\rangle_{10011}
                 \right).
\end{eqnarray}
{We} use the same labeling {for} the annihilation operators {and frequency modes}. 
{Note} that the authors of the experiment \cite{Danan2013} {set} the phase shift in the 
inner interferometer to $\varphi = 0,\,\pi$ [see Fig.2 in Ref.~\cite{Danan2013}].

{To explain the results of} \cite{Danan2013}, it is crucial to 
correctly describe the post-selection process caused by photon detection and subsequent 
frequency mode analysis. 
When one particular frequency mode is post-selected, the 
information about the photon being {in a superposition of} 
other frequency modes is erased. {The} frequency modes {are} 
orthogonal{, i.e., one} can perform a direct 
deterministic signal frequency {analysis to}
{distinguish between the modes}~\cite{Barnett2009}.
{Thus,} post-selecting a specific frequency mode 
makes the information about {the} other frequency modes unavailable.

{Post-selection is a well established technique in quantum state engineering used, e.g.,
in optimal quantum cloning \cite{QSE}.} Here, 
the post-selection on the photon {that interacted with}
mirror A {is formally equivalent} to the projection of the output state onto the state
{\begin{equation}\label{eq:projA}
  |\Pi_A\rangle = \sum_{A,B,C,E,F=0,1} \delta_{A,1}|1\rangle_{ABCEF},
\end{equation}
where $\delta_{A,1}$ is Kronecker's delta. Post-selection on any other
mode is also defined by Eq~(\ref{eq:projA}), where A in Kronecker's
delta is replaced with a chosen mode, i.e.,  $A\leftrightarrow X$ for $X=B,\,C,\,E,\,F$.} 
{Now,} one can immediately see from {Eq.}~(\ref{eq:finalFock})
{that}
\begin{equation}\label{eq:propA}
  |\langle\Pi_A |\psi_\mathrm{out}\rangle|^2 = \frac{1}{9}
\end{equation}
{and, similarly,}
\begin{equation}\label{eq:propBC}
  |\langle\Pi_B |\psi_\mathrm{out}\rangle|^2 =  
  |\langle\Pi_C |\psi_\mathrm{out}\rangle|^2 = \frac{1}{9}.
\end{equation}
{The structure of the output state implies that post-selecting 
modes E and F yields the same results for both. For mode
E one obtains}
\begin{eqnarray}\nonumber
 \langle\Pi_E |\psi_\mathrm{out}\rangle  & = &\frac{1}{3}\left(       
    -\langle\Pi_E |1\rangle_{01011} 
    +\mathrm{e}^{\mathrm{i}\varphi}
       \langle\Pi_E |1\rangle_{10011}
  \right)
  \\\label{eq:EF}
  && = {\frac{2}{3}\mathrm{e}^{\mathrm{i}\varphi/2}\sin\varphi = \langle\Pi_F |\psi_\mathrm{out}\rangle .}
\end{eqnarray}
Thus, for $\varphi = 0$ none of the frequency modes marked as {E  or F} will contribute to the final state. 
{If the phase shift is set to $\varphi = \pi$,} 
both modes will appear as $|\langle\Pi_E |\psi_\mathrm{out}\rangle|^2 = |\langle\Pi_F |\psi_\mathrm{out}\rangle|^2 = \frac{4}{9}$.


{
Let us use the  theoretical framework established above
to explain the experimental data in Ref.~\cite{Danan2013}. }
In case of constructive interference [$\varphi=\pi$, see Fig.~\ref{fig}(a)], all 
frequencies $f_X$ for $X=A,\,B,\,C,\,E,\,F$ are present in the power spectrum {recorded by} detector D. Intensities {for} 
frequencies $f_E$ and $f_F$ are four times higher than others because of constructive 
{interference} in {the} inner MZI. 
Destructive interference appears for $\varphi=0$ [see Fig.~\ref{fig}(b)] and 
it removes the peaks for frequencies $f_E$ and $f_F$ from the power spectrum. 
The peaks for frequencies $f_A$ and  $f_B$ remain constant because {post-selecting}
on mode A or B is equivalent to post-selecting on a photon traveling via the post-selected 
arm of the inner MZI, so there is no interference at the BS$_3$.

The experimental results presented {in Figs.~2(a) and 2(b) of Ref.}~\cite{Danan2013} 
{agree with our theoretical predictions.}
{If the photon reflected from mirror} C is blocked 
and the phase shift is $\varphi=0$ [see Fig.~\ref{fig}(c)], 
our theoretical prediction does not match the experimental data
from Ref.~\cite{Danan2013}. In this case we predict
that $f_A,\,f_B$ should be constant (similarly to the previously 
discussed cases). This is because it is possible to  distinguish
between photons reflected from mirrors A and B. When mode 
post-selection (power spectrum analysis) 
is performed the interference on BS$_3$ is effectively 
removed. Therefore, it should be possible to observe
the intensity peaks for $f_A$ and $f_B$. As we present below, 
the intensity peaks for $f_A$ and $f_B$ are also
predicted by the classical theory of light.

The classical approach {to} deriving frequency-mode amplitudes is based on 
{the} standard electromagnetic-wave theory (Supplemental Material of Ref.~\cite{Danan2013}).
We repeated this classical procedure, but we did not {keep track of} the normalization factors
and took only those parts of the {expressions} that were relevant. The amplitude of the electric field 
at the detector D 
takes in general the form of
\begin{eqnarray*}
  \Psi(y,t) &\propto& \kappa\mathrm{e}^{-(y-d_C)^2}
  - \mathrm{e}^{-(y-d_A -d_E -d_F)^2}\\
  &&
  + \mathrm{e}^{\mathrm{i}\varphi}\mathrm{e}^{-(y-d_B -d_E -d_F)^2},
\end{eqnarray*}
where $\kappa = 1$ ($\kappa = 0$) when mode $c$ is open (closed) and  $d_X$ are small shifts in direction  $y$ oscillating with frequencies of the relevant mirror labeled with $X = A,B,C,E,F$. The amplitude expressed using the paraxial approximation reads
\begin{eqnarray}
\label{eq:amplitude}
\Psi(y,t) &\propto & \mathrm{e}^{-y^2}\left[\kappa-1+\mathrm{e}^{\mathrm{i}\varphi} + 2\kappa yd_C - 2yd_A + \mathrm{e}^{\mathrm{i}\varphi}2yd_B\right. \nonumber\\
&&  \left.+ 2y\left(\mathrm{e}^{\mathrm{i}\varphi}-1\right)\left(d_E+d_F\right)\right]
\end{eqnarray}

Let us use Eq.~(\ref{eq:amplitude}) to calculate the intensities in all three scenarios shown in  Fig.~2 of \cite{Danan2013}. In the first scenario (a) there is constructive interference in the small MZI and mode $c$ is open, i.e., we set $\kappa=1$ and $\varphi=\pi$. The associated  field amplitude reads
\begin{eqnarray}\nonumber
\Psi_{\mathrm{a}}(y,t) &\propto& \mathrm{e}^{-y^2}\left[1 + 2y\left(d_A + d_B - d_C  +2d_E+2d_F\right)\right]\\
\label{eq:amplitude_a}
\end{eqnarray}
The measurement performed by Danan \etal \cite{Danan2013} consists of evaluating the power spectrum of the function
\begin{eqnarray}
\label{eq:dI}
\Delta I_a(t) \equiv \int_0^\infty |\Psi_a(y,t)|^2 \mathrm{d}y - \int_{-\infty}^0 |\Psi_a(y,t)|^2 \mathrm{d}y.
\end{eqnarray}
The  Fourier transform of  $\Delta I_a(t) $ is
$
\Delta I_a(f) \propto \delta(f-f_A) + \delta(f-f_B) +\delta(f-f_C) +2\delta(f-f_E) +2\delta(f-f_F).
$
It provides five peaks in the associated power spectrum $|\Delta I_a(f)|^2$, where peaks corresponding 
to frequencies $f_A,\, f_B,\, f_C$  have four times smaller area than peaks for $f_E,\, f_F$.
Note that the corresponding intensity difference ratios are 1:2 (in contrast to our quantum model, which describes 
a sum of intensities). 

In the second scenario (b) mode $c$ is open ($\kappa =1$) and there is destructive interference in the small MZI ($\varphi=0$). Thus, the field amplitude 
\begin{eqnarray}
\label{eq:amplitude_b}
\Psi_b(y,t) &\propto& \mathrm{e}^{-y^2}\left[1 + 2y\left(d_C - d_A + d_B \right)\right]
\end{eqnarray}
provides the power spectrum of $|\Delta I_a(f)|^2$ containing three balanced peaks associated with mirrors A, B and C (the same result is provided by our quantum model).

In the third scenario (c), the mode $c$ is blocked ($\kappa =0$) and $\varphi=0$. The amplitude 
\begin{eqnarray}
\label{eq:amplitude_c}
\Psi_c(y,t) &\propto& 2y\mathrm{e}^{-y^2}\left(d_A - d_B \right)
\end{eqnarray}
in this case provides intensity $|\Psi_c(y,t)|^2$ that is even function of $y$, hence $\Delta I_c=0$. Therefore, no peaks are observed by Danan \etal \cite{Danan2013}. On the other hand, it follows from Eq. (\ref{eq:amplitude_c}) that the amplitude and the resulting intensity oscillates with frequencies $f_A$ and $f_B$. It means that these quantities 
include the information  about the photons impinging on the mirrors A and B. However, the specific quantity measured by Danan \etal \cite{Danan2013} ignores this information. This leads us to conclude that $\Delta I$ is not a reliable which-path witness because it ignores some of the available information.

In order to use the available information to its fullest extent we propose to use spectrum of the overall intensity
$$
I_T(f) \equiv \int_{-\infty}^\infty |\Psi(y,f)|^2 \mathrm{d}y
$$
where $\Psi(y,f)$ is a Fourier transform of the field $\Psi(y,t)$ (the parameters of which can be established from the setup configuration and $I(t)$ measurements). This quantity is a more reliable which-path witness and it corresponds to our quantum model. Note that $I_T$ does not vanish for even $|\Psi(y,f)|^2$ (in contrast to $\Delta I$). The total intensity $I_T(f)$ contains contributions from the respective mirror frequencies. The weights associated with the specific mirrors can be calculated as $w(f_X) = I_T(f_X)\,\mathrm{d}f,$, where $f_X$ stands for the respective mirrors frequencies. This approach produces exactly the same results as obtained by the previously mentioned quantum approach. In scenario (a), the intensity $I_T(f)$ provides the weights $w(f_X)$ of the five peaks that have the 1:4 ratio. In scenario (c) $I_T(f)$ does not hide the which-path information encoded in the presence of frequency peaks corresponding to mirrors A and B (see Fig. \ref{fig}).

In this paper, we  described the spectra at the output port of 
nested MZI with vibrating mirrors applying both quantum
and classical theories of light. The quantum approach employed 
the standard formalism of annihilation operators
and one-state vectors in Fock's basis.  
The time-dependent transverse profiles of beams were described
classically and used in the analysis of the spectrum 
of the electric field. Using the classical approach, we have explained the results observed by Danan \etal \cite{Danan2013} and established that the quantity they used is not a reliable which-path witness. We, therefore, propose to acquire spectrum of the overall intensity instead and show that it produces well expected results that can be described both classically and in a quantum way.

\section*{Acknowledgements}
The authors thank Jára Cimrman for his helpful suggestions and David Schmid for a very fruitful discussion.
K.~L. acknowledges support by the Czech Science Foundation (Grant No.
13-31000P). D.~J. acknowledges support by project IGA\_PrF\_2014005 of IGA UP Olomouc. 
K.~B. acknowledges support by the Foundation for Polish Science and
the Polish National Science Centre under grant No.
DEC-2013/11/D/ST2/02638. Finally, the authors acknowledge the project No. LO1305 of the
Ministry of Education, Youth and Sports of the Czech Republic.


\end{document}